\documentclass[pdflatex,sn-mathphys-ay]{sn-jnl}

\usepackage{graphicx}%
\usepackage{multirow}%
\usepackage{amsmath,amssymb,amsfonts}%
\usepackage{amsthm}%
\usepackage{mathrsfs}%
\usepackage[title]{appendix}%
\usepackage{xcolor}%
\usepackage{textcomp}%
\usepackage{manyfoot}%
\usepackage{booktabs}%
\usepackage{algorithm}%
\usepackage{algorithmicx}%
\usepackage{algpseudocode}%
\usepackage{listings}%
\usepackage{natbib}
\usepackage{mdframed}
\usepackage{array}
\usepackage{textcomp}

\theoremstyle{thmstyleone}%
\theoremstyle{thmstyletwo}%

\theoremstyle{thmstylethree}%

\definecolor{lightergray}{rgb}{0.9, 0.93, 0.96}

\begin{document}

\title[Article Title]{Leveraging Graph-RAG and Prompt Engineering to Enhance LLM-Based Automated Requirement Traceability and Compliance Checks}


\author*[1]{\fnm{Arsalan} \sur{Masoudifard}}\email{arsalan.masoudi@sharif.edu}

\author[1]{\fnm{Mohammad} \sur{Mowlavi Sorond}}\email{mmowlavi2002@gmail.com}

\author[1]{\fnm{Moein} \sur{Madadi}}\email{moeinmadadi2002@gmail.com}

\author[2]{\fnm{Mohammad} \sur{Sabokrou}}\email{mohammad.sabokrou@oist.jp}

\author[1]{\fnm{Elahe} \sur{Habibi}}\email{e.habibi1367@gmail.com}

\affil*[1]{\orgdiv{Department of Computer Engineering}, \orgname{Sharif University of Technology}, \orgaddress{ \city{Tehran},  \country{Iran}}}

\affil*[2]{ \orgname{Okinawa Institute of Science and Technology}, \orgaddress{ \city{Okinawa},  \country{Japan}}}


\abstract{

Ensuring that Software Requirements Specifications (SRS) align with higher-level organizational or national requirements is vital, particularly in regulated environments such as finance and aerospace. In these domains, maintaining consistency, adhering to regulatory frameworks, minimizing errors, and meeting critical expectations are essential for the reliable functioning of systems. The widespread adoption of large language models (LLMs) highlights their immense potential, yet there remains considerable scope for improvement in retrieving relevant information and enhancing reasoning capabilities. This study demonstrates that integrating a robust Graph-RAG framework with advanced prompt engineering techniques, such as Chain of Thought and Tree of Thought, can significantly enhance performance. Compared to baseline RAG methods and simple prompting strategies, this approach delivers more accurate and context-aware results. While this method demonstrates significant improvements in performance, it comes with challenges. It is both costly and more complex to implement across diverse contexts, requiring careful adaptation to specific scenarios. Additionally, its effectiveness heavily relies on having complete and accurate input data, which may not always be readily available, posing further limitations to its scalability and practicality.
}




\maketitle

\section{Introduction}\label{sec1}

Early identification of risks in the Software Development Life Cycle (SDLC), particularly during the requirements specification phase, is critical to the success of software development \citep{AbuSalim2020}. Addressing issues such as requirement changes and compliance issues at this stage is essential to prevent their propagation and mitigate costly challenges later in the project \citep{Verner2014, Pacheco2018, Bibhash2016}. Specifically, software in regulated environments demands meticulous attention during the requirements specification phase to ensure adherence to stringent standards and regulatory frameworks, which, if overlooked, can lead to significant failures and project setbacks \citep{Marques2019}.

Recent advancements in Natural Language Processing (NLP), particularly in the development of Large Language Models (LLMs), have motivated researchers in requirements engineering to explore the potential of these tools in enhancing Software Requirement Specification (SRS) documents. Effective compliance checking in SRS documents plays a crucial role in mitigating many associated risks. Luitel et al. \citeyearpar{Luitel2024} demonstrate how BERT, a Large Language Model, is employed to detect and address incomplete requirements by predicting missing terminology, thereby improving the overall completeness of SRS. In addition to enhancing completeness, LLMs have also shown promise in ensuring regulatory compliance. For instance, the study by Hassani et al. \citeyearpar{hassani2024} illustrates how Data Processing Agreements (DPAs) can be evaluated for compliance with the General Data Protection Regulation (GDPR), a legal framework aimed at ensuring data privacy in the European Union, using advanced techniques powered by Large Language Models. Their work highlights how automating compliance checks can streamline the validation process, reducing manual efforts and improving accuracy in legal and regulatory adherence.

However, LLMs face significant challenges when validating SRS, particularly in regulated environments such as finance and aerospace, where adherence to strict regulations is critical. These challenges include difficulties in maintaining context across extensive documents, which can result in incomplete or inaccurate analysis during requirement compliance checks. A potential solution is to provide the most relevant reference text, allowing for more precise validation of whether a requirement is being violated \citep{Spoletini2024}. Moreover, hallucination, where models generate factually incorrect yet plausible content, remains a critical issue, particularly when retrieval mechanisms are insufficient and reasoning capabilities fail.  Additionally, scalability and performance constraints limit their effectiveness in large, complex projects, and inherent biases in LLMs can result in prioritizing common problems over more nuanced, domain-specific issues \citep{Huang2023}.

In this work, we propose an automated framework to address the challenges of validating SRS in regulated environments. The framework leverages Graph-RAG to retrieve the most relevant content from reference texts. Graph-RAG enhances retrieval accuracy by constructing a graph-based text index from regulatory articles, enabling the identification of closely related entities and clusters for more precise compliance checking \citep{Fan2024, Gao2023}. Once the relevant content is retrieved, the framework verifies whether the requirements adhere to these references using advanced prompting techniques. Methods such as Chain of Thought (CoT) and Tree of Thought (ToT) are incorporated to enhance reasoning capabilities. CoT provides a transparent, step-by-step analysis of complex queries, while ToT explores multiple reasoning paths in parallel, enabling a more comprehensive evaluation of potential solutions \citep{Wei2022, Yao2023}. This combination of advanced retrieval and reasoning techniques ensures a robust and accurate approach to requirement validation.

We tested this framework in two regulated environments. The first involves the SRS of an online broker application in Iran, which must adhere to stock market standards defined in \citep{IranStockArticles}. The second is the SRS for NASA’s X-38 Fault Tolerant System Services, which requires compliance with fault-tolerant systems and redundancy standards defined by its Fault Tolerant Parallel Processor (FTPP) architecture \citep{X38SRS2000}.

\begin{figure}[htbp]
  \centering
  \includegraphics[width=\linewidth]{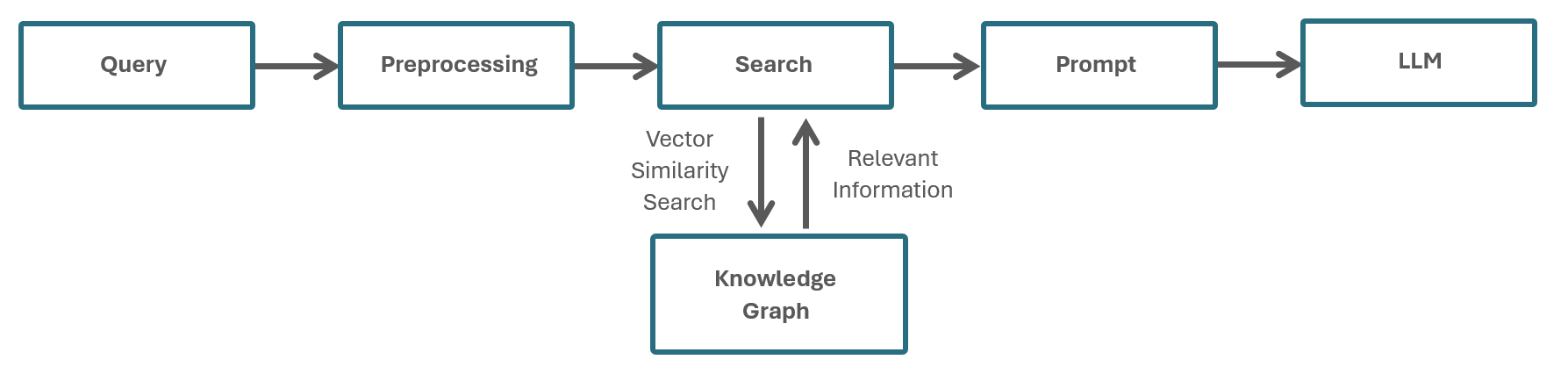}
  \caption{The simplified workflow for checking requirement compliance.}
  \label{fig:fig_1}
\end{figure}

As shown in Figure~\ref{fig:fig_1}, the automated framework begins by receiving a requirement as a query to retrieve relevant content from the reference text. The next step is preprocessing, where the requirement is cleaned and divided into smaller, manageable components to enhance downstream processing. The Search module then interacts with a Knowledge Graph constructed from higher-level requirement documents. The retrieved information is passed to the Prompt stage, where it is structured into a query to check for potential violations in the SRS. Both the Search and Prompt modules involve multiple steps, which are detailed in the following sections. Finally, the LLM (GPT-4o) evaluates the prompt to determine if the requirement violates the retrieved regulations.

The main contributions of this paper are as follows:

\begin{itemize}
    \item \textbf{Automated Compliance Check Framework:} We present an improved framework using Graph-RAG for precise retrieval and advanced prompting techniques to enhance reasoning and accuracy, outperforming baseline methods in detecting SRS violations of higher-level requirements.

    \item \textbf{Comparison of Graph-RAG and Baseline RAG:} We compare Graph-RAG with baseline RAG to demonstrate improvements in retrieval accuracy and compliance check.
    
    \item \textbf{Effectiveness of Advanced Prompting Techniques:} We explore how Tree of Thoughts (ToT) and Chain of Thought (CoT) enhance reasoning and interpretability in compliance checks.
    
    \item \textbf{Addressing Limitations and Risks of Over-reliance on AI:} We address AI over-reliance risks and propose strategies to improve robustness and trustworthiness.
\end{itemize}

The remainder of this paper is organized as follows: Section 2 provides the necessary background, outlining key concepts and technologies relevant to this study. Section 3 reviews related works. Section 4 presents the proposed methodology, detailing the framework design, Graph-RAG integration, and prompting techniques. Section 5 evaluates the framework's performance through experiments and comparative analysis. Section 6 discusses the results, identifies limitations, and explores the implications of this work. Finally, Section 7 concludes the paper and outlines potential directions for future research.

\section{Background}\label{sec2}

In this section, we introduce and define the fundamental concepts and methodologies that form the basis of this work, including compliance check, Large Language Models, Retrieval-Augmented Generation, and prompt engineering.

\subsection{Compliance Check}
Traceability is a fundamental quality attribute of an SRS as defined in \citep{IEEE830-1998}. It refers to the ability to link each requirement to its origin, such as stakeholder needs or higher-level specifications, and to corresponding design, implementation, and testing elements. Traceability ensures that requirements are consistently aligned with their source and remain verifiable throughout the development process. Since our work focuses on cross-referencing, we primarily emphasize linking requirements to higher-level specifications. This approach not only reinforces traceability but also enhances impact analysis, validation, and verification by maintaining clear and structured connections between requirements and their origins. The importance of traceability lies in its role in reducing ambiguity, managing changes effectively, and ensuring that the final product meets all specified needs and standards.

\subsection{Large Language Models}

LLMs are advanced AI systems, typically built on the Transformer architecture, with billions of parameters trained on massive text corpora. These models demonstrate emergent abilities, such as in-context learning and step-by-step reasoning, making them highly effective at solving complex tasks, including language understanding and text generation \citep{Zhao2024}. Their Transformer-based architecture, as exemplified by models like GPT and BERT, allows them to process large-scale textual data and generate context-aware responses with high accuracy. These capabilities make LLMs particularly useful for cross-referencing, as they can analyze documents, understand intricate linguistic patterns, and generate reasoning steps for validation, aligning seamlessly with automated frameworks for regulatory adherence explored in this work \citep{Raiaan2024}.

\subsection{Retrieval Augmented Generation}

RAG is a hybrid framework that combines the generative capabilities of LLMs with retrieval mechanisms to enhance performance on knowledge-intensive tasks. Unlike traditional LLMs, which rely solely on parametric memory, RAG introduces a non-parametric memory in the form of an external database or document index. By retrieving relevant information during the generation process, RAG ensures that outputs are more accurate, contextually grounded, and up-to-date \citep{Lewis2020}.

Building upon the RAG framework, Graph-RAG enhances retrieval capabilities by constructing a graph-based text index from external knowledge sources. This method leverages the interconnected nature of data, representing entities and their relationships as a graph structure. By doing so, Graph-RAG improves the precision of retrieval by enabling contextualized exploration of related entities and concepts. This graph-based representation is particularly effective for tasks requiring fine-grained reasoning, such as cross-referencing, where requirements often reference interconnected standards, regulations, and supporting documentation. Graph-RAG not only retrieves relevant content but also provides structured clusters of related entities, facilitating comprehensive cross-referencing and traceability analysis \citep{Gao2023}.

\subsection{Prompt Engineering}

Prompt engineering is the process of designing and refining input instructions, or "prompts," to guide LLMs toward generating desired outputs. A well-crafted prompt provides context, instructions, and sometimes examples to elicit specific responses, improving the quality, relevance, and accuracy of the generated results \citep{Marvin2024}. 

Chain of Thought (CoT) and Tree of Thought (ToT) are advanced prompting methods designed to enhance reasoning capabilities in LLMs. CoT prompting involves breaking down a problem into a sequence of intermediate reasoning steps, enabling the model to handle complex tasks like arithmetic, commonsense, and symbolic reasoning. This step-by-step decomposition not only allows for better allocation of computation resources but also offers interpretability, as the reasoning process becomes transparent and debuggable \citep{Wei2022}.

ToT expands on CoT by framing problem-solving as a search process through a tree structure, where each node represents a partial solution and branches signify different reasoning paths. Unlike CoT, which sequentially generates reasoning steps, ToT allows LLMs to explore multiple reasoning paths simultaneously, evaluate their progress, and backtrack when necessary. This deliberate search mechanism significantly improves performance on tasks requiring exploration and strategic planning, such as mathematical puzzles and creative writing \citep{Yao2023}.

\section{Related Works}\label{sec3}

In recent years, significant research has focused on the application of LLMs and NLP techniques in Software Requirements Engineering (SRE). Arora et al. \citeyearpar{Arora2023} provide a comprehensive exploration of LLMs' potential in areas such as elicitation, specification, analysis, and validation. Similarly, Jin et al. \citeyearpar{Jin2024} discuss the challenges associated with LLMs in SRE, including the potential for bias in generated content, hallucinations, and the necessity of human supervision to ensure that generated requirements are accurate and useful. Necula et al. \citeyearpar{Necula2024}  explore how NLP enhances various aspects of requirements management, such as elicitation, specification, and validation, by reducing ambiguities and improving accuracy. Additionally, the paper highlights both the benefits and challenges of incorporating machine learning and AI-driven techniques into SRE workflows.

Recent research in LLMs also delves into their potential in the field of software engineering and compliance checking. Hassani et al. \citeyearpar{hassani2024} discuss the challenges and opportunities of using LLMs for formal compliance analysis. Their approach considers broader contexts than individual sentences and even provides explanations and rationales for compliance decisions, which previous approaches lacked.  Similarly, Cejas et al. \citeyearpar{Amaral2023} propose an NLP-based automated approach for verifying compliance of Data Processing Agreements (DPAs) with GDPR. Their framework uses semantic role labeling and phrasal-level representations to identify violations and recommend corrective measures, achieving significant precision and recall improvements compared to baseline NLP techniques. This approach underscores the potential of integrating NLP with domain expertise for enhancing compliance checking processes. In the context of finance, Berger et al. \citeyearpar{Berger2023} investigated the challenges of automating financial statements in comparison with regulatory frameworks. They analyzed the efficiency of both proprietary and open-source LLMs, including LAMA-2 and GPT-4, in detecting noncompliance cases. And about standard compliance Arora et al. \citeyearpar{Arora2024} propose the CompliAT framework, which utilizes LLMs to enhance the compliance of assistive technology product specifications with international standards, focusing on tasks such as terminology consistency, product classification, and compliance checking.

Focusing on software requirements, Luitel et al. \citeyearpar{Luitel2024} propose enhancing the completeness of natural language requirements by employing BERT, a specific type of LLM, to detect potential incompleteness. Similarly, Fantechi et al. \citeyearpar{Fantechi2023} examine the capability of GPT-3.5 in identifying inconsistencies within natural language software requirements, comparing its performance to that of human experts. Complementing this line of research, Gärtner and Göhlich \citeyearpar{Gartner2024} introduce ALICE, a system combining formal logic with LLMs to detect contradictions in controlled natural language requirements. Their approach employs a decision tree model and LLM-based prompts to identify contradictions, achieving higher accuracy and recall rates than LLM-only methods, thereby demonstrating its potential for improving consistency in complex specification documents.

Several studies have explored the application of LLMs in generating SRS documents. Krishna et al. \citeyearpar{Krishna2024} evaluated the performance of LLMs in generating, validating, and correcting SRS documents, comparing their results to human benchmarks. In another study, Ronanki et al. \citeyearpar{Ronanki2023} assessed the quality of LLM-generated requirements across multiple attributes, including abstraction, atomicity, and consistency. Additionally, Yeow et al. \citeyearpar{Yeow2024} focused on the ability of LLMs to generate survey and interview questions for requirement gathering, evaluating the clarity and relevance of the content.

In the context of using RAG and LLMs, Zhang and colleagues \citeyearpar{Zhang2023} conducted an empirical evaluation, demonstrating the effectiveness of generative large language models in handling requirements information retrieval tasks. Their study highlights the models' accuracy in extracting domain-specific terms and features from software artifacts, even in zero-shot settings, where the model performs without prior task-specific training. Similarly, Edwards and co-authors \citeyearpar{Edwards2023} investigate a hybrid approach combining context retrieval, LLMs, and knowledge graphs to enhance text generation, with applications across various domains, including SRS documentation. Sami et al. \citeyearpar{Sami2024} explore the integration of RAG techniques with large language models to enhance the prioritization of software requirements. Their approach leverages the strengths of RAG to efficiently generate and rank user stories, aligning development efforts with business objectives while addressing the inherent challenges of traditional prioritization methods.

\section{Automated Method}\label{sec4}
Our automated approach is designed with two primary objectives: first, to identify relevant content from regulations and higher-level requirements that align with our requirements using Graph-RAG; and second, to enhance the reasoning capabilities of LLMs through advanced prompting techniques to improve the detection of wrong information. In this section, we provide an explanation of the dataset, describe the construction of Graph-RAG, outline the search process, and detail the implementation of CoT and ToT prompting methods.

\subsection{Dataset Preparation}\label{subsec4}
Our dataset comprises two SRS documents along with supplementary materials, including regulatory articles, standards, and higher-level requirements, which the SRS documents are expected to trace and adhere to.

\subsubsection{Requirement Document}\label{subsubsec4}

The first requirement document in our dataset is the SRS for a broker application, which must comply with higher-level specifications outlined in regulatory articles defined by the national stock organization in Iran. The SRS is structured according to the IEEE Std 830-1998 standard \citeyearpar{IEEE830-1998}, ensuring that each requirement is systematically detailed. Table \ref{tab1} provides an example of a structured requirement from this document. It outlines the key components of a requirement, including its title, description, inputs, processing steps, outputs, and error-handling mechanisms. For instance, the "Bank Account Validation" requirement describes how the system validates customer bank account information against an internal database, ensuring compliance with regulatory expectations while addressing potential error scenarios. This document includes 40 functional requirements and 66 non-functional requirements.

\begin{table}[h]
\caption{An example of a structured requirement from SRS\_Broker.}\label{tab1}%
\begin{tabular}{@{}lp{8cm}@{}}
\toprule
\textbf{Component} & \textbf{Explanation} \\
\midrule
Title & Bank Account Validation \\
\midrule
Description   & Validate the bank account entered by the customer.    \\
\midrule
Input    & Account number, bank name    \\
\midrule
Processing    & The bank account information is checked against an internal database. If the account matches a record in the internal database, it is considered valid.     \\
\midrule
Output    & Confirmation of account validity or invalidity.     \\
\midrule
Error Handling    & If there is a communication issue with the bank interface, the system proceeds without notifying the customer. \\
\botrule
\end{tabular}
\end{table}

Unlike the structured SRS\_Broker, the second document, SRS\_Aero, lacks a structured format defined by IEEE 830 and contains software requirements written as single statements, as referenced in \citep{X38SRS2000}. These requirements are designed to be traceable to the X-38 Fault Tolerant Parallel Processor (FTPP) Requirements document, which was prepared for NASA's Johnson Space Center. Identified as 297749 Rev F, this document outlines the software and external interface requirements for the X-38 project, a crew return vehicle developed for the International Space Station. In highly regulated environments like aerospace, cross-cheking plays a critical role in ensuring traceability and alignment of requirements with overarching specifications and standards. This approach helps maintain compliance, facilitates validation, and reduces the risk of oversight in complex systems. Table \ref{tab2} provides examples of unstructured requirements from SRS\_Aero, illustrating how individual requirements are expressed in this document.

\begin{table}[h]
\caption{Examples of requirements in unstructured requirements: SRS\_Aero.}\label{tab2}%
\begin{tabular}{@{}lp{8cm}@{}}
\toprule
\textbf{Requirement ID} & \textbf{Specification} \\
\midrule
194  & Whenever a power-on reset occurs, System Initialization shall perform the following functions. 
\\
\midrule
234    & As part of System Initialization, the Boot ROM shall ] be configured to, after completing IBIT, call the manufacturer-supplied VxWorks Board Support Package (BSP) initialization software followed by a call to the FTSS System Initialization software.  \\
\midrule
014  &  System Initialization shall initiate the watchdog timer.    \\
\midrule
292    & System Initialization shall enable and reset the processor’s watchdog timer such that, in the absence of a fault, the watchdog timer does not expire and reset the 
processor.\\
\midrule
008    & System Initialization shall synchronize the FCP virtual group in the presence of a power on skew of 2.5 seconds. \\

\botrule
\end{tabular}
\end{table}

\subsubsection{Regulatory Documents and Standards}\label{subsubsec4}
For the SRS\_Broker document, mandatory requirements must conform to higher-level regulations outlined in \citep{IranStockArticles}, which include 30 articles specifically addressing online brokers. Additionally, adherence to the guidance of standards from \citep{ISO27001} is encouraged. The SRS was manually refined by two software engineers working on the actual project, resulting in a total of 223 pairs of requirements and related articles. As shown in Table \ref{tab3}, the label "irrelevant" indicates requirements for which no corresponding content could be identified through cross-referencing. The labels "compliant" and "non-compliant" represent traceable requirements, where "compliant" signifies conformance with the referenced materials, while "non-compliant" indicates that the requirements do not adhere to the expected standards or regulations after tracing. This distribution provides valuable insight into the alignment of the SRS\_Broker document with higher-level regulatory and best-practice frameworks.

A total of 105 software requirements were selected from \citep{X38SRS2000}, representing the unstructured requirements in the SRS\_Aero document, which are intended to be traced to the FTPP specifications. Among these, 20 requirements were found to lack sufficient relevant information for cross-referencing, rendering their traceability infeasible. A total of 346 pairs of requirements in the SRS and FTPP specifications were identified as compliant, based on the ground truth document provided by the original authors. Additionally, 54 requirements were intentionally manipulated to be non-compliant, primarily by removing or replacing critical informational words with incorrect details. This dataset underscores the challenges posed by differing contexts and formats of the requirements, which complicate traceability and conformance validation processes.

\begin{table}[h]
\caption{Ground truth class distribution across SRS\_Broker and SRS\_Aero.}\label{tab3}%
\begin{tabular}{@{}lcc@{}}
\toprule
\textbf{Label}     & \textbf{SRS\_Broker} & \textbf{SRS\_Aero} \\
\midrule
Compliant           & 121                              & 346           \\
Non-Compliant      & 68                              & 54           \\
Irrelevant          & 34                              & 20           \\
\midrule
Total               & 223                             & 420           \\
\botrule
\end{tabular}
\end{table}

\subsection{Preprocessing}\label{subsec4} 

\begin{itemize}
    \item \textbf{Text Cleaning:} In the preprocessing phase, text cleaning is performed to remove noise and irrelevant content, such as stop words, special characters, and formatting artifacts. This step ensures that the textual data is concise, structured, and ready for further processing. By eliminating distractions that do not contribute to meaningful relationships, text cleaning improves the quality of subsequent operations, including chunking and graph construction. This process is particularly important for handling unstructured documents, such as SRS\_Aero, where raw text often contains extraneous elements that can interfere with accurate graph representation.
\\
    \item \textbf{Extracting the Glossary:} The glossary of terms is extracted as part of preprocessing. Feeding the glossary into the graph is crucial because it provides a predefined set of domain-specific terms and definitions that can enhance the semantic richness of the graph. Incorporating the glossary ensures that key terms are properly represented as entities within the graph, improving the traceability and contextual understanding of requirements in relation to higher-level standards and regulations.
\\
    \item \textbf{Chunking Documents:} The final step before constructing the graph is to segment cross-referenced documents into manageable text chunks. Chunking is essential to balance the efficiency and accuracy of downstream processes. The size of these chunks significantly impacts the performance of subsequent tasks. Smaller chunks provide better recall of entities and relationships as they fit within the context window of LLMs, minimizing the risk of losing critical information. However, smaller chunks require more processing calls to cover the entire document. Conversely, larger chunks reduce the number of processing calls but may degrade recall due to the limited context-handling capacity of LLMs. Thus, the choice of chunk size represents a trade-off between processing efficiency and the quality of extracted information, making it a crucial consideration in the Graph-RAG pipeline.
\end{itemize}

\begin{figure}[htbp]
  \centering
  \includegraphics[width=\linewidth]{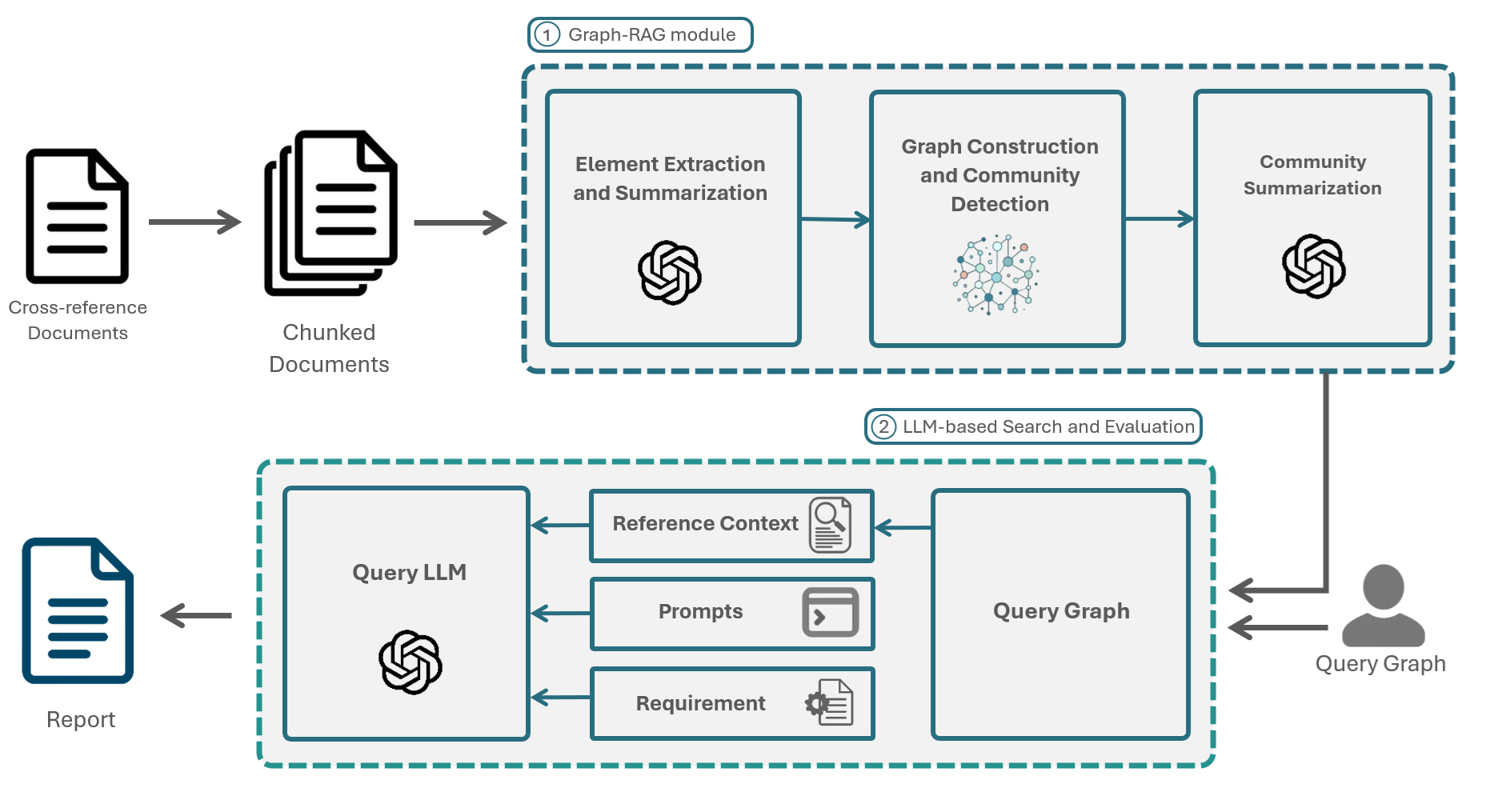}
  \caption{The automated framework diagram.}
  \label{fig:fig_2}
\end{figure}

\subsection{Graph Construction}\label{subsec4}

After chunking the documents, as illustrated in Figure \ref{fig:fig_2}, the next step involves constructing the graph. The graph construction method is based on the approach proposed by \citep{Edge2024}. However, to align with the specific requirements and characteristics of our context, modifications to the original method are necessary.

\subsubsection{Element Extraction and Summarization}

    The first step in this process involves identifying and extracting entities (nodes) and relationships (edges) from each text chunk. This is accomplished using domain-specific prompts tailored for the LLM to recognize relevant elements within the text. To ensure comprehensive extraction, especially for larger chunk sizes, iterative rounds of extraction (gleanings) are performed. These rounds refine the results by addressing any overlooked entities or relationships, thereby enhancing the accuracy and completeness of the extracted elements. This method leverages LLMs, invoking the model with the general prompt structure shown in Figure \ref{fig:Entity_prompt} for each text chunk to extract domain-specific entities and relations.

\begin{figure}[h!]
\begin{mdframed}[linewidth=1pt, backgroundcolor=lightergray]
   \textbf{Entity Extraction Prompt Overview:} 
\small

\begin{itemize}
    \item \textbf{Goal:} Given a text document relevant to software requirements and a list of pre-defined entity types (\texttt{article}, \texttt{standard}, \texttt{requirement}), identify entities of those types and the relationships among them.
    \item \textbf{Steps:}
    \begin{enumerate}
        \item Extract all entities, including:
        \begin{itemize}
            \item \textbf{entity\_name:} Name of the entity (capitalized).
            \item \textbf{entity\_type:} One of the pre-defined types (\texttt{article}, \texttt{standard}, \texttt{requirement}).
            \item \textbf{entity\_description:} A detailed explanation of the entity's attributes and relevance. \\
            Format: (entity, entity\_name, entity\_type, entity\_description)
        \end{itemize}
        \item Identify all related entity pairs (source\_entity, target\_entity), including:
        \begin{itemize}
            \item \textbf{relationship\_description:} Explanation of the relationship in the context of software requirements.
            \item \textbf{relationship\_strength:} Numeric score indicating the strength of the relationship. \\
            Format: (relationship, source\_entity, target\_entity, relationship\_description, relationship\_strength)
        \end{itemize}
        \item Return a single list of all entities and relationships, separated by commas. End the output with a clear delimiter indicating completion.
        \item Examples for few-shot learning will be here.
    \end{enumerate}
\end{itemize}

\end{mdframed}
\caption{Entity Extraction Prompt Overview.}
\label{fig:Entity_prompt}
\end{figure}

The entity extraction prompt is designed to identify specific entities (\texttt{article}, \texttt{standard}, \texttt{requirement}) and their relationships within a text document relevant to software requirements. This prompt operates as a few-shot prompt, a technique where a limited number of examples are provided within the prompt to guide the model’s understanding of the task. By including examples of entities and relationships in the input, the model leverages these patterns to generalize and accurately identify similar structures in the target document. This approach enhances the prompt’s effectiveness, especially for tasks requiring domain-specific knowledge and structured outputs.

    Once the entities and relationships are extracted, they are summarized into concise, descriptive blocks of text for each graph element. This summarization step leverages LLM-generated abstractive summaries to provide semantically rich and consistent representations of the extracted elements. These summaries ensure that the graph maintains clarity and coherence while facilitating downstream tasks such as community detection, query-focused summarization, and knowledge retrieval. 

\subsubsection{Graph Construction and Community Detection}

After extracting entities and their relationships, a homogeneous, undirected, and weighted graph is constructed. In this graph, nodes represent the entities, while edges capture the relationships between them, with weights indicating the strength or relevance of these relationships. The next step involves detecting communities within the graph. A community is a group of nodes that are more densely connected to each other than to the rest of the graph, representing clusters of related entities. Identifying these communities helps to organize the graph into semantically meaningful clusters, enabling more efficient and context-aware analysis, such as generating targeted summaries or answering specific queries.

\begin{figure}[htbp]
  \centering
  \includegraphics[width=\linewidth]{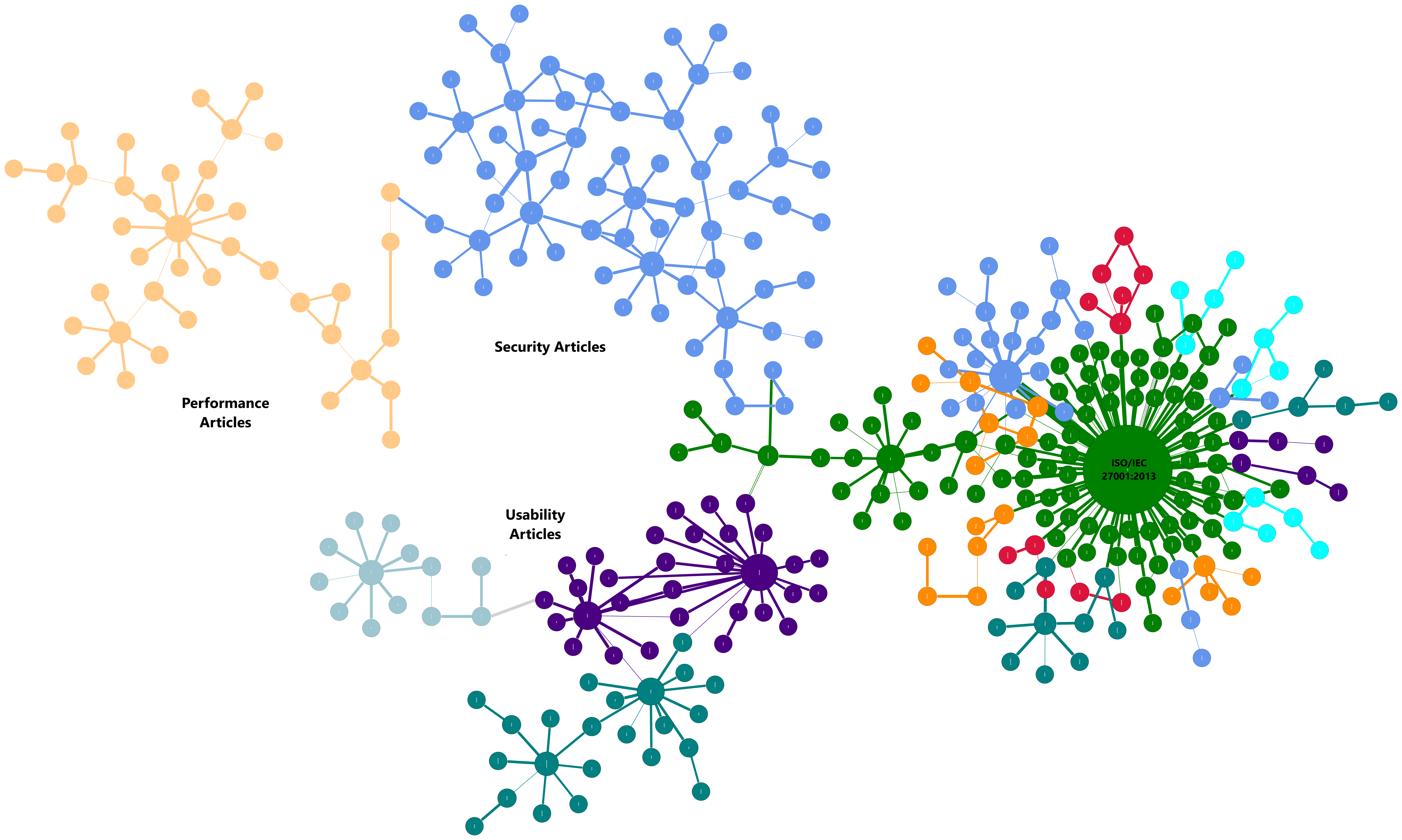}
  \caption{Holistic view of communities in the constructed graph.}
  \label{fig:fig_3}
\end{figure}

Sub-community ranking prioritizes the most relevant sub-communities within a larger community for summarization, particularly under token constraints or other context limitations. Initially, sub-communities were ranked using static metrics, such as the number of connections or predefined importance based on entity type. While effective in identifying prominent entities, this static approach often overlooked smaller sub-communities containing valuable, context-specific information. To address this, we introduced two key modifications. First, a dynamic weighting system now adjusts rankings based on the terms in the query, ensuring that smaller but highly relevant sub-communities are prioritized. Second, a similarity threshold evaluates how closely related sub-communities are to the query, including those with indirect but thematically relevant information.
Figure \ref{fig:fig_3} presents a comprehensive visualization of the communities detected in the graph, each represented by distinct clusters of nodes and edges. Key clusters include "Articles," "Security Requirements," and "Broker Application Requirements," showcasing interconnected entities and relationships within their respective domains.

\subsubsection{Community Summarization}
Community summarization is a critical step in the Graph-RAG framework as it provides a concise and structured understanding of the entities and relationships within each cluster. By summarizing communities, the framework organizes complex and interconnected data into meaningful segments, making it easier for users to navigate and extract insights. These summaries enable efficient query processing, as each community can independently contribute partial answers that are later aggregated into a comprehensive global response. This approach improves the scalability of the framework for large datasets while ensuring that users gain a clear and organized view of the data, highlighting key themes and relationships within each community \citep{Edge2024}.

\begin{figure}[h!]
\begin{mdframed}[linewidth=1pt, backgroundcolor=lightergray]
   \textbf{Community Summarization Prompt Overview:} 
\small

\begin{itemize}
    \item \textbf{Goal:} Generate a report on a software requirements community, summarizing its key entities, relationships, compliance with standards, technical implications, articles, and potential impact on software development. The report informs decision-makers about the community's relevance and significance.

    \item \textbf{Report Structure:}
    
       The report should include the following sections:
    \begin{itemize}
        \item \textbf{TITLE:} A short, specific name representing the community's key entities, including named entities where possible.
        \item \textbf{SUMMARY:} An overview of the community's structure, entity relationships, and significant information related to software requirements.
        \item \textbf{IMPACT SEVERITY RATING:} A 0-10 score reflecting the community's importance in software development and requirements engineering.
        \item \textbf{RATING EXPLANATION:} A brief explanation of the impact severity rating.
        \item \textbf{DETAILED FINDINGS:} 5-10 key insights, each with:
        \begin{itemize}
            \item A concise summary.
            \item Detailed explanatory text grounded in predefined rules, focusing on software requirements engineering.
        \end{itemize}
    \end{itemize}
    \item Output format will be defined here.

    \end{itemize}

    \end{mdframed}
    \caption{Community summarization prompt overview.}
    \label{fig:Community_Prompt}
\end{figure}

This section is also LLM-based and requires a carefully designed prompt to guide the model. The prompt, illustrated in Figure \ref{fig:Community_Prompt}, facilitates the generation of a comprehensive report summarizing a software requirements community. The report includes key sections such as the community's title, a summary of its structure and relationships, an impact severity rating (0-10) with an explanation, and detailed findings highlighting significant insights. This structured output helps decision-makers assess the community's compliance with standards, technical implications, articles, and noteworthy claims, ensuring informed decision-making in software development projects.

\subsection{Search}\label{subsec4}
The search process in Graph-RAG involves querying the knowledge graph constructed from the input documents. The graph consists of nodes (entities) and edges (relationships) enriched with metadata such as rank, weight, and direction. When a query is made, the system traverses the graph to retrieve the most relevant nodes and their connections, using the metadata to prioritize the results. 

\begin{figure}[htbp]
  \centering
  \includegraphics[width=\linewidth]{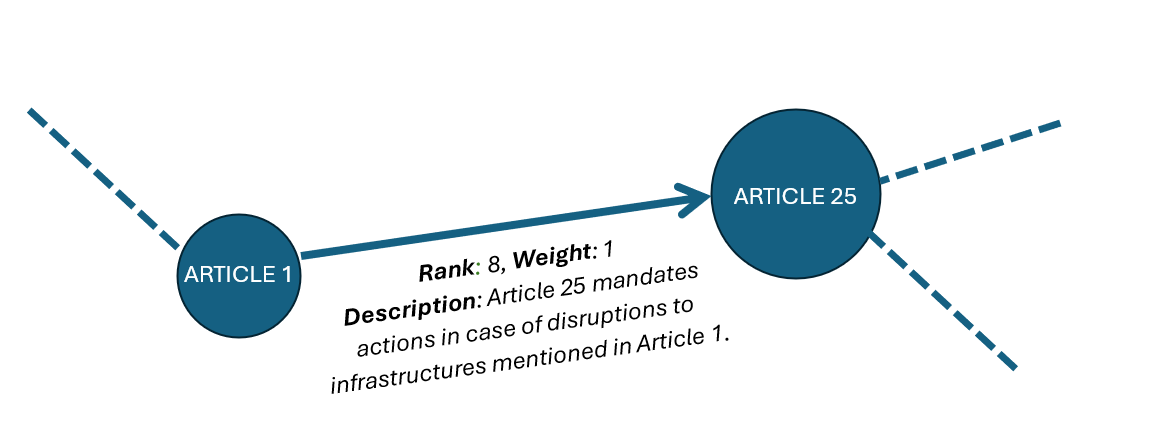}
  \caption{Visualization of a relationship between two entities in the graph.}
  \label{fig:two_nodes}
\end{figure}

Figure \ref{fig:two_nodes} illustrates the connection between two entities, ARTICLE 1 and ARTICLE 25, both related to SRS\_Broker. The directed arrow represents a relationship where ARTICLE 25 provides additional mandates or requirements connected to the context of ARTICLE 1.

\begin{itemize}
    \item \textbf{Rank:}  Indicates the prominence or importance of a node (entity) in the graph, often based on its connectivity or degree within the network. Higher-ranked entities are prioritized during search, as they are more likely to be central to the query's context.
    \item \textbf{Weight:}  Represents the strength or relevance of the relationship between two nodes. Edges with higher weights are given precedence in the traversal process, ensuring that stronger and more meaningful connections are considered in the results.
    \item \textbf{Direction:} Direction: Captures the flow of influence or dependency between entities. In directed graphs, the edge direction can determine whether a particular relationship is relevant to the query's context. For example, it can help distinguish between a requirement referencing a standard versus a standard influencing a requirement.
\end{itemize}

With the graph constructed, it can now be queried to retrieve the most relevant content from cross-referenced documents. A key factor in this process is the similarity threshold, which determines the level of contextual alignment required between the query and the graph elements. Lowering the similarity threshold allows for the retrieval of more content, providing broader context for later evaluation. The impact of adjusting this threshold on the quality and relevance of the retrieved content will be analyzed in the evaluation section.

\subsection{Designing Prompts}\label{subsec4}

As illustrated in Figure \ref{fig:fig_2}, following the search step, the reference context is retrieved to assess whether the requirement aligns with it. The effectiveness of this evaluation depends significantly on the design of the prompts, which play a critical role in guiding the reasoning process. Further details on the prompt structures are provided in the appendix.

\begin{itemize}
\item \textbf{Input/Output Prompt:} The simplest approach, IO prompting, involves assessing whether a requirement aligns with its reference text by determining if it is fully addressed and identifying any inconsistencies or violations. The outcome is categorized as either "Conforms" or "Violates," accompanied by a concise explanation of the reasoning. This straightforward method ensures clarity and precision in evaluating alignment.

\item \textbf{Chain-of-Thought Prompt:} CoT prompting leverages a structured, multi-step process to evaluate the alignment between a requirement and its reference text. The process begins with the first step, which decomposes both the requirement and reference text into three formal logical components: Purpose (objective or goal), Action (specific actions or processes), and Conditions/Constraints (requirements for the actions to occur). This breakdown establishes a clear foundation for further analysis.

In the second step, each of these components is compared across the requirement and reference text. The comparison involves assessing whether the Purpose aligns, partially aligns, or conflicts, identifying inconsistencies or overlaps in the Action, and evaluating the consistency or gaps in the Conditions/Constraints. For each component, detailed reasoning is provided to highlight alignment or conflict, ensuring a comprehensive evaluation.

The final step synthesizes the component-level analyses to deliver an overall assessment. This includes determining whether the requirement conforms to or violates the reference text, supported by a clear rationale that highlights the key areas of alignment, inconsistencies, or conflicts. This step-by-step logical approach, encoded into the prompts, ensures a rigorous evaluation while maintaining clarity and focus on critical details.

\item \textbf{Tree-of-Thought Prompt:} ToT prompting applies a multi-agent reasoning framework to assess the alignment between a requirement and its reference text. The process begins with three independent agents breaking down the requirement and reference text into logical components: Purpose, Action, and Conditions/Constraints. Each agent provides an individual analysis, ensuring diverse perspectives without cross-influence. A final arbiter consolidates their outputs, selecting the most accurate and comprehensive breakdown.

In the second step, the agents analyze the consolidated breakdown to determine how well each component aligns, partially aligns, or conflicts between the requirement and the reference text. Each agent provides detailed reasoning for their evaluation, emphasizing different aspects of alignment. The arbiter then synthesizes these analyses, identifying the strongest reasoning to produce a consolidated evaluation for each component.

In the final step, the agents collectively assess whether the requirement overall "Conforms" or "Violates" the reference text. Their evaluations consider all components—Purpose Alignment, Action Consistency, and Conditions/Constraints Alignment—and are supported by detailed rationales. The arbiter reviews the agents' assessments to produce a final decision, either adopting one perspective or synthesizing them for a balanced conclusion. This iterative, multi-agent approach ensures a robust and comprehensive evaluation of alignment and conformance.
\end{itemize}

\section{Evaluation}\label{sec5}
In this section, we begin by evaluating the search capabilities of Graph-RAG in comparison to the baseline RAG. Subsequently, we assess its reasoning ability to determine the alignment between the retrieved reference content and the given requirement.

This baseline RAG implementation combines dense retrieval with a generative language model to deliver context-aware answers \citep{Lewis2020}. It begins by embedding documents into dense vector representations using a pre-trained SentenceTransformer model (all-MiniLM-L6-v2). These embeddings enable efficient retrieval of relevant content. When a query is input, the system computes its embedding and identifies the most similar documents from the indexed collection. The retrieved documents are then formatted into a prompt, providing context for the generative model (GPT-4o and GPT-4o-mini) to answer the query. This approach ensures that responses are informed by the retrieved context, making it effective for tasks requiring direct access to relevant information. However, it does not account for complex relationships or structured dependencies between entities, serving as a baseline for comparison with more advanced systems like Graph-RAG.

\subsection{Retrieving Reference Text}\label{subsec5}

The first step in our process involves querying the graph by providing a requirement and retrieving relevant content from the reference text. As illustrated in Figure \ref{fig:graph_query}, we prompt the graph with a requirement from SRS\_Aero, which returns related content based on a calculated similarity score. While content with high similarity scores is often sufficient, there are cases where lowering the similarity threshold is necessary to ensure that the intended content is retrieved. Table \ref{tab:similarity_scores} presents examples of reference text content that should have been retrieved from the graph, along with their corresponding similarity scores. These scores indicate the degree of alignment between the requirement and the reference text, influenced by factors such as lexical overlap, semantic connections, and contextual nuances.

\begin{figure}[h!]
\begin{mdframed}[linewidth=1pt, backgroundcolor=lightergray]
   \textbf{Graph Query:} \\
   
\small
Retrieve content that is directly related to the following requirement: \\
\textbf{SRS008:} System Initialization shall synchronize the FCP virtual group in the presence of a power on skew of 3.5 seconds.

    \end{mdframed}
    \caption{Graph query prompt.}
    \label{fig:graph_query}
\end{figure}

The similarity scores between the SRS and the reference texts are primarily influenced by shared keywords, semantic alignment, and the presence of explicit conditions. Keywords such as "FCP," "synchronize," "power-on skew," and "initialization" establish a strong lexical connection, helping the model recognize related content. The semantic logic, which involves alignment in intent (e.g., achieving synchronization under timing constraints), further reinforces the relationship. Additionally, reference texts that explicitly mention timing constraints, such as "2.5-second power-on skew" or "synchronization errors," provide closer matches, resulting in higher similarity scores. Texts with broader or tangential relevance, like descriptions of control flow, score lower due to weaker alignment with the specific requirement.

\begin{table}[h]
\caption{Similarity scores between SRS requirement and reference texts.}\label{tab:similarity_scores}%
\begin{tabular}{@{}p{10cm}>{\centering\arraybackslash}p{3cm}@{}}
\toprule
\textbf{Reference Text} & \textbf{Similarity Score} \\
\midrule
In the presence of a maximum 2.5-second power-on skew, the FTPP system shall (3.1.6) be capable of completing FCC system power-up and initialization without synchronization errors. & 0.92 \\
\midrule
Start Up shall (3.3.2.12) synchronize its FCP with other operational FCPs. & 0.85 \\
\midrule
Start Up shall (3.3.2.18) be able to synchronize all operational FCPs in the presence of this skew in the power-on sequence. & 0.95 \\
\midrule
Start Up shall (3.3.2.19) test to ensure that all four FCPs are synchronized. & 0.78 \\
\midrule
Control flow of the four FCPs shall (3.3.11.2) be similar, if not identical. & 0.70 \\
\botrule
\end{tabular}
\end{table}

In this evaluation, we compare four models: RAG and Graph-RAG, each implemented with both GPT-4o and GPT-4o-mini. Given that constructing a knowledge graph using LLMs can incur significant computational costs, incorporating a more cost-effective variant such as GPT-4o-mini offers valuable insights. By analyzing the performance of GPT-4o-mini, we can assess the trade-offs between cost efficiency and model effectiveness, enabling a more practical understanding of Graph-RAG’s scalability and applicability across varying computational resources. The coverage is evaluated automatically based on exact matches.

\begin{figure}[htbp]
  \centering
  \includegraphics[width=\linewidth]{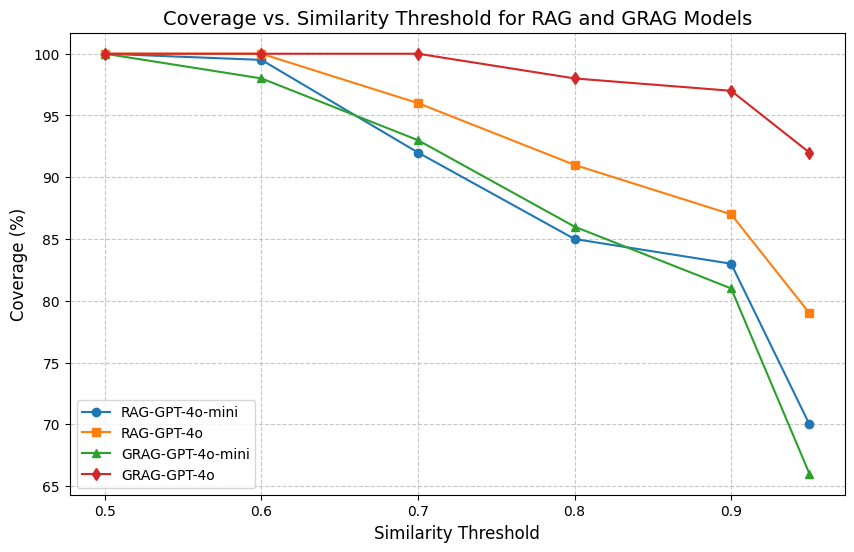}
  \caption{Coverage of ground-truth reference texts under varying similarity thresholds.}
  \label{fig:similarity_graph}
\end{figure}

Figure \ref{fig:similarity_graph} illustrates the coverage percentage achieved by four models—RAG-GPT-4o-mini, RAG-GPT-4o, GRAG-GPT-4o-mini, and GRAG-GPT-4o—across varying similarity thresholds. The X-axis represents the similarity threshold, ranging from 0.5 to 0.95, while the Y-axis indicates the coverage percentage. The coverage is calculated across both datasets, comprising a total of 206 requirements. As shown in the figure, focusing solely on contexts with higher similarity thresholds results in the loss of significant reference content.

As shown, GRAG-GPT-4o consistently achieves the highest coverage across all thresholds, reflecting its superior ability to retrieve relevant content, even at higher thresholds. RAG-GPT-4o-mini and RAG-GPT-4o demonstrate a gradual decline in coverage as the threshold increases, highlighting the trade-off between precision and recall in retrieval. GRAG-GPT-4o-mini, though slightly less effective than its full-scale counterpart, still outperforms the RAG models at higher thresholds. This comparison underscores the impact of incorporating graph-based retrieval techniques and advanced language models on improving retrieval accuracy.

\subsection{Results}\label{subsec5}

A similarity threshold of 0.7 is chosen because a higher threshold causes excessive information loss, while a lower threshold introduces issues with context and precision. Using this threshold, reference text is retrieved, and conformance is checked across four models and three prompts. The evaluation metrics used are Precision, Recall, and F1-Score. Since non-conformance manipulations involve word misplacement or masking, the evaluation process is automated. The four models are a combination of our two RAG models and two LLMs, GPT-4o and GPT-4o-mini.

\begin{figure}[htbp]
  \centering
  \includegraphics[width=\linewidth]{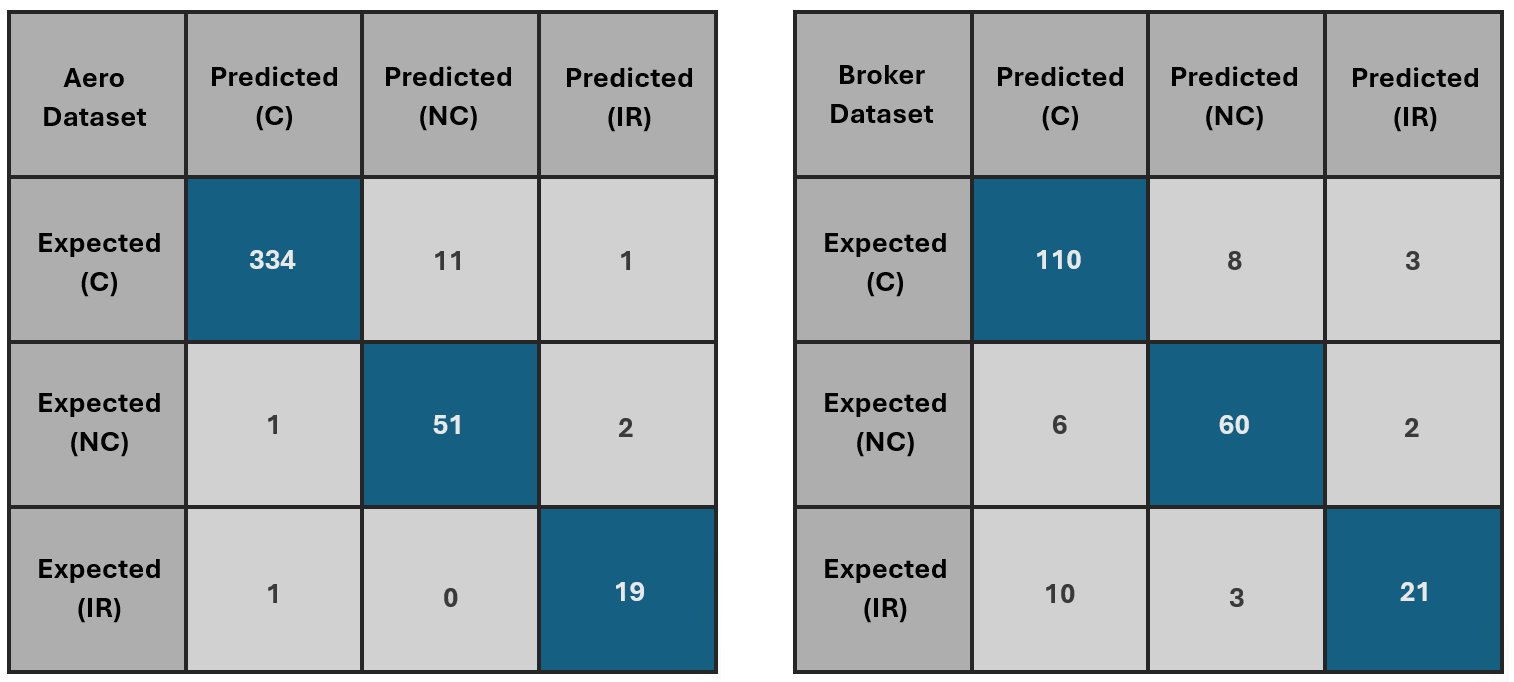}
  \caption{Confusion matrices of GRAG-GPT-4o in ToT prompting for both datasets.}
  \label{fig:confusion_matrices}
\end{figure}

As shown in Figure \ref{fig:confusion_matrices}, the confusion matrices illustrate the classification performance across the three categories: C for Conform, NC for Non-Conform, and IR for Irrelevant. Each matrix visualizes the true and predicted labels, with diagonal elements representing correct classifications. Among the datasets, SRS\_Aero demonstrates superior performance despite being larger in size. This is likely due to its requirements being more concise and exhibiting greater similarity in semantic style, which aids in achieving better classification outcomes.

In Table \ref{tab4}, we evaluate the models and prompts based on their ability to identify non-conformed requirements, using metrics such as Precision (P), Recall (R), and F1-Score (F1). The results highlight the strengths of different prompting techniques and models. The Tree of Thoughts (ToT) prompt generally achieves better reasoning capabilities, leading to higher precision. However, the weaker language model the GPT-4o-mini tends to classify more requirements as non-conformed, resulting in higher recall but lower F1-Score due to a lack of precision. For example, under the ToT prompt, RAG-GPT-4o-mini achieves a recall of 83.33\% but an F1-Score of 64.52\%, while GRAG-GPT-4o demonstrates its strength with the highest F1-Score of 86.33\% due to balanced precision (84.51\%) and recall (88.24\%). These results emphasize the trade-offs between precision, recall, and model capability when evaluating non-conformed requirements.

\begin{table}[htbp]
\caption{Results of non-compliant pairs in the SRS\_Broker dataset.}\label{tab4}
\centering
\begin{tabular}{@{}lccc|ccc|ccc@{}}
\toprule
\textbf{Label} & \multicolumn{3}{c|}{\textbf{IO}} & \multicolumn{3}{c|}{\textbf{CoT}} & \multicolumn{3}{c}{\textbf{ToT}} \\
               & \textbf{P} & \textbf{R} & \textbf{F1} & \textbf{P} & \textbf{R} & \textbf{F1} & \textbf{P} & \textbf{R} & \textbf{F1} \\
\midrule
RAG-GPT-4o-mini  & 50.43 & 80.56 & 62.03 & 51.30 & 81.94 & 63.10 & 52.63 & 83.33 & 64.52 \\
RAG-GPT-4o       & 60.23 & 73.61 & 66.25 & 59.09 & 72.22 & 65.00 & 62.22 & 77.78 & 69.14 \\
\midrule
GRAG-GPT-4o-mini & 62.77 & 90.77 & 74.21 & 64.89 & 93.85 & 76.73 & 67.37 & 94.12 & 78.53 \\
GRAG-GPT-4o      & 76.71 & 83.58 & 80.00 & 79.73 & 86.76 & 83.10 & 84.51 & 88.24 & \textbf{86.33} \\
\botrule
\end{tabular}
\end{table}

Table \ref{tab5} presents the results for the SRS\_Aero dataset, where the ToT prompting method combined with GRAG-GPT-4o achieved the highest F1-score, outperforming other methods. However, the differences between models are less pronounced in this dataset compared to the previous ones considering it is more imbalanced. This observation emphasizes the importance of input data characteristics, as the nature of the dataset significantly impacts the performance variation among models and prompts.

\begin{table}[htbp]
\caption{Results of non-compliant pairs in the SRS\_Aero dataset.}\label{tab5}
\centering
\begin{tabular}{@{}lccc|ccc|ccc@{}}
\toprule
\textbf{Label} & \multicolumn{3}{c|}{\textbf{IO}} & \multicolumn{3}{c|}{\textbf{CoT}} & \multicolumn{3}{c}{\textbf{ToT}} \\
               & \textbf{P} & \textbf{R} & \textbf{F1} & \textbf{P} & \textbf{R} & \textbf{F1} & \textbf{P} & \textbf{R} & \textbf{F1} \\
\midrule
RAG-GPT-4o-mini          & 45.05  & 75.93  & 56.55  & 51.16  & 78.57  & 61.97  & 53.49  & 85.19  & 65.71 \\
RAG-GPT-4o           & 57.14  & 74.07  & 64.52  & 66.15  & 76.79  & 71.07  & 69.23  & 83.33  & 75.63 \\
\midrule
GRAG-GPT-4o-mini           & 50.51  & 92.59  & 65.36  & 55.43  & 94.44  & 69.86  & 56.67  & 94.44  & 70.83 \\

GRAG-GPT-4o          & 60.52  & 85.19  & 70.77  & 73.53  & 92.59  & 81.97  & 82.26  & 94.44  & \textbf{87.93} \\
\botrule
\end{tabular}
\end{table}

Returning to the contributions of this research, we demonstrated that employing various RAG techniques is crucial for identifying the most relevant content for our requirements. For detecting violations of higher-level requirements, we utilized different prompting methods, which revealed that results can vary significantly depending on the chosen approach. This highlights the need for careful and deliberate use of LLMs and AI-generated content, particularly in critical fields where errors could result in serious consequences, emphasizing the risks of over-reliance on such technologies.

\section{Discussion}\label{sec6}
This section begins by outlining the limitations in this automated framework, addressing aspects that may impact its functionality or applicability. Following this, we discuss potential threats to validity, highlighting factors that could influence the reliability or generalizability of the framework's results.

\subsection{Limits}\label{subsec6}

The effectiveness of Graph-RAG systems heavily relies on the quality and completeness of the input data. Inaccurate or missing information can lead to suboptimal knowledge graphs, adversely affecting performance \citep{Peng2024}. Peng et al. \citeyearpar{Peng2024} also mentioned that To mitigate these issues, it is recommended to conduct regular data audits, establish data governance protocols, and automate data cleaning processes to ensure that the data feeding into Graph-RAG systems is clean, structured, and relevant.

Graph-RAG systems introduce several complexities due to their reliance on structured relational knowledge and dynamic data handling. Unlike traditional RAG methods, Graph-RAG must accurately retrieve and utilize interconnected data elements, such as nodes, triples, paths, or subgraphs, which adds layers of complexity in terms of both processing and relevance. Additionally, the static nature of many current graph databases limits the system's ability to adapt to real-time updates, making dynamic integration a challenging task. The choice of retrieval granularity further complicates the process, requiring a careful balance between retrieval efficiency and maintaining contextual relevance. Moreover, the need to compress extensive graph-based contexts without losing critical information presents another significant hurdle, especially when working with large-scale or complex data. These challenges highlight the intricate interplay of retrieval, representation, and integration in Graph-RAG systems, necessitating advanced techniques for optimization and scalability \citep{Peng2024}.

Constructing and retrieving information from Graph-RAG involves numerous LLM queries, which can be costly. Additionally, our results indicate that using a less powerful model, such as GPT-4o-mini, is less effective compared to GPT-4o when benchmarked against baseline RAG systems.

Adhering to higher-level requirements or regulations is crucial for ensuring compliance and maintaining the integrity of software systems. While AI models and automated approaches can significantly enhance efficiency and accuracy in analyzing and implementing these requirements, over-reliance on them poses inherent risks. Automated systems may overlook nuanced interpretations or context-specific elements of regulations, leading to potential non-compliance or misalignment with critical standards. This underscores the importance of incorporating human oversight and validation to complement AI-driven processes, ensuring a balanced approach that leverages the strengths of automation while mitigating its limitations.

\subsection{Threats to Validity}\label{subsec6}

Internal Validity: The internal validity of this study could be influenced by the manual modification of requirements to create non-compliance with reference materials. However, the search process remains unaffected, as both the requirements and reference materials are derived from authentic sources, ensuring the integrity of the retrieval aspect.

External Validity: The external validity is limited by the domain-specific nature of the study. Since the framework was tested in regulated environments like finance and aerospace, its applicability to other domains with different compliance requirements remains uncertain. However, the results obtained from both datasets were not contradictory, reinforcing the reliability of the framework within these tested domains. Broader testing across diverse domains is necessary to assess its generalizability.

Cost-Effectiveness Threats: The high computational cost associated with methods like Graph-RAG and Tree of Thoughts presents a threat to scalability. Implementing such resource-intensive techniques may not be practical for large-scale or resource-constrained applications, requiring further optimization to enhance efficiency.

\section{Conclusion and future work}\label{sec7}

This study demonstrated the application of advanced retrieval-augmented generation techniques and large language models to enhance compliance checking in software requirements specifications. By integrating Graph-RAG with LLMs, the approach effectively extracts concise and relevant information from higher-level requirements that SRS documents must comply with. To further enhance reasoning and detection accuracy, various prompting strategies, including Chain of Thought and Tree of Thought, were employed, highlighting the significant impact of prompt design on model performance.

The results indicate that while weaker LLMs like GPT-4o-mini are more cost-effective, they exhibit limitations in reasoning capabilities. These models often adopt a more holistic perspective, leading to the identification of more violations of higher-level requirements, albeit with reduced precision. On the other hand, Graph-RAG combined with the ToT prompting strategy outperformed other methods, showcasing its ability to deliver more accurate and context-aware results. However, the effectiveness of this approach depends heavily on the meticulous design of the graph construction framework, including the inner prompts and the careful manipulation of thresholds, which require effort to optimize for specific contexts. This underscores the trade-offs between model capability, implementation complexity, and cost-effectiveness.

Additionally, the effectiveness of this automated framework heavily depends on the quality and completeness of input data. Over-reliance on AI-driven systems poses risks, as these systems may overlook nuanced or context-specific regulatory elements. Complementing automation with human oversight is essential to account for these subtleties, ensuring a balanced and reliable compliance framework that leverages the strengths of AI while mitigating its limitations.

For future work, the framework could be extended to accommodate the diverse forms and structures of SRS documents, emphasizing the need for generalization across different domains and formats. Developing robust search tools to automate the retrieval of relevant reference texts would further enhance the system's efficiency and applicability. These advancements would ensure broader adaptability and streamline the process of compliance verification in varied contexts, ultimately making the framework more scalable and versatile.

\backmatter












\bigskip





\bibliography{sn-article}
\clearpage 
\begin{appendices}

\section{IO Prompt}\label{secA1}

\begin{figure}[h!]
\begin{mdframed}[linewidth=1pt, backgroundcolor=lightergray]

\textbf{User Prompt (IO):}  
\begin{itemize}
    \item \textbf{Requirement:} \texttt{\{requirement\}}
    \item \textbf{Reference Text:} \texttt{\{reference\_text\}}
\end{itemize}

\noindent Assess whether the requirement aligns with the reference text by following these steps:
\begin{enumerate}
    \item Identify if the requirement is fully addressed in the reference text.
    \item Determine if there are any inconsistencies or violations in the requirement with respect to the reference text.
    \item If the requirement conforms, state \texttt{Conforms}.
    \item If the requirement violates or contradicts the reference text, state \texttt{Violates} and briefly explain why.
\end{enumerate}

\noindent Output your assessment in the following format:
\begin{itemize}
    \item \textbf{Assessment:} [Conforms | Violates]
    \item \textbf{Explanation:} [Provide a concise explanation of your assessment.]
\end{itemize}

\end{mdframed}
\caption{IO prompt.}
\label{fig:condition_example}
\end {figure}

\begin{figure}[h!]
\begin{mdframed}[linewidth=1pt, backgroundcolor=lightergray]
\textbf{User Prompt (First CoT):}  
\begin{itemize}
    \item \textbf{Requirement:} \texttt{\{requirement\}}
    \item \textbf{Reference Text:} \texttt{\{reference\_text\}}
\end{itemize}

\noindent Break down the components of the requirement and the reference text into the following formal logical structure:
\begin{enumerate}
    \item \textbf{Purpose:} What is the objective or goal of the requirement or reference text?
    \item \textbf{Action:} What specific actions or processes are described?
    \item \textbf{Conditions/Constraints:} What conditions or constraints must be satisfied for the actions to occur?
\end{enumerate}

\noindent Output the results in this format:
\begin{itemize}
    \item \textbf{Requirement Components:}
    \begin{itemize}
        \item \textbf{Purpose:} [Requirement's purpose]
        \item \textbf{Action:} [Requirement's action]
        \item \textbf{Conditions/Constraints:} [Requirement's conditions/constraints]
    \end{itemize}
    \item \textbf{Reference Text Components:}
    \begin{itemize}
        \item \textbf{Purpose:} [Reference text's purpose]
        \item \textbf{Action:} [Reference text's action]
        \item \textbf{Conditions/Constraints:} [Reference text's conditions/constraints]
    \end{itemize}
\end{itemize}

\end{mdframed}
\caption{First CoT prompt.}
\label{fig:condition_example}
\end {figure}

\begin{figure}[h!]
\begin{mdframed}[linewidth=1pt, backgroundcolor=lightergray]
\textbf{User Prompt (Second CoT):}  
\textbf{The following are the extracted components of the requirement and the reference text:}
\begin{itemize}
    \item \textbf{Requirement Components:}
    \begin{itemize}
        \item Purpose: \texttt{\{Requirement\_Purpose\}}
        \item Action: \texttt{\{Requirement\_Action\}}
        \item Conditions/Constraints: \texttt{\{Requirement\_Conditions\_Constraints\}}
    \end{itemize}
    \item \textbf{Reference Text Components:}
    \begin{itemize}
        \item Purpose: \texttt{\{Reference\_Text\_Purpose\}}
        \item Action: \texttt{\{Reference\_Text\_Action\}}
        \item Conditions/Constraints: \texttt{\{Reference\_Text\_Conditions\_Constraints\}}
    \end{itemize}
\end{itemize}
\end{mdframed}
\caption{Second CoT prompt.}
\label{fig:condition_example}
\end {figure}

\begin{figure}[h!]
\begin{mdframed}[linewidth=1pt, backgroundcolor=lightergray]
\textbf{User Prompt (Third CoT):}  
\noindent Compare the corresponding components (\textbf{Purpose vs. Purpose, Action vs. Action, Conditions/Constraints vs. Conditions/Constraints}) and provide an analysis for each:

\begin{enumerate}
    \item \textbf{Purpose Comparison:} Describe whether the purposes align, partially align, or conflict. Highlight any key differences or overlaps.
    \item \textbf{Action Comparison:} Identify whether the actions align, partially align, or conflict. Explain any inconsistencies or missing elements.
    \item \textbf{Conditions/Constraints Comparison:} Assess whether the conditions or constraints are consistent. Note any contradictions, gaps, or additional requirements.
\end{enumerate}

\noindent \textbf{Output your analysis in the following format:}
\begin{itemize}
    \item \textbf{Purpose Analysis:} [Analysis of alignment or conflict]
    \item \textbf{Action Analysis:} [Analysis of alignment or conflict]
    \item \textbf{Conditions/Constraints Analysis:} [Analysis of alignment or conflict]
\end{itemize}
\end{mdframed}
\caption{Third CoT prompt.}
\label{fig:condition_example}
\end {figure}

\begin{figure}[h!]
\begin{mdframed}[linewidth=1pt, backgroundcolor=lightergray]

\begin{small}
\textbf{User Prompt (First ToT):}  \\
\noindent You are a set of three reasoning agents (\textbf{Agent A}, \textbf{Agent B}, \textbf{Agent C}), followed by a final \textbf{Arbiter}. Your goal is to break down the given requirement and the reference text into logical components.

\noindent \textbf{Instructions for Agents:}

    Each agent should work independently and not share or alter their reasoning based on the others. Each agent should provide a breakdown for both the requirement and the reference text into: \{Purpose, Action, Condition/Constraints\}

\noindent \textbf{Given Input:} \{Requirement, Reference\_Text\}

\noindent \textbf{Format for Each Agent's Response:}
\begin{itemize}
    \item \textbf{Agent [Name] Analysis:}
    \begin{itemize}
        \item \textbf{For both Requirement and Reference Components :}
        \begin{itemize}
            \item Purpose: ...
            \item Action: ...
            \item Conditions/Constraints: ...
        \end{itemize}
    \end{itemize}
\end{itemize}

\noindent \textbf{Steps:}
\begin{enumerate}
    \item \textbf{Agent A:} Provide your breakdown.
    \item \textbf{Agent B:} Provide your breakdown.
    \item \textbf{Agent C:} Provide your breakdown.
\end{enumerate}

\noindent After all agents have provided their breakdowns:

\noindent \textbf{Arbiter Instructions:}
\begin{itemize}
    \item Review all three agents' analyses.
    \item Compare and identify the most accurate or comprehensive elements.
    \item Produce a final, consolidated breakdown that selects the best \textbf{Purpose}, \textbf{Action}, and \textbf{Conditions/Constraints} for both the requirement and the reference text.
\end{itemize}

\noindent \textbf{Final Output Format (Arbiter's Consolidated Answer):}
\begin{itemize}
    \item \textbf{Final Consolidated Components:}
    \begin{itemize}
        \item \textbf{Requirement and Reference Components:}
        \begin{itemize}
            \item Purpose: [Consolidated Purpose]
            \item Action: [Consolidated Action]
            \item Conditions/Constraints: [Consolidated Conditions/Constraints]
        \end{itemize}
    \end{itemize}
\end{itemize}
\end{small}

\end{mdframed}
\caption{First ToT prompt.}
\label{fig:condition_example}
\end {figure}

\begin{figure}[h!]
\begin{mdframed}[linewidth=1pt, backgroundcolor=lightergray]
\textbf{User Prompt (Second ToT):}  \\
\noindent You are a set of three reasoning agents (\textbf{Agent A}, \textbf{Agent B}, \textbf{Agent C}), followed by a final \textbf{Arbiter}. You have received a final consolidated breakdown of the requirement and reference text from the previous step.

\noindent \textbf{Consolidated Breakdown:} \texttt{\{final\_consolidated\_breakdown\}}

\noindent \textbf{Instructions for Agents:}
\begin{itemize}
    \item Each agent should independently analyze how the \textbf{Purpose}, \textbf{Action}, and \textbf{Conditions/Constraints} of the requirement compare to those of the reference text.
    \item Identify if each component aligns, partially aligns, or conflicts.
    \item Provide reasoning for each comparison.
\end{itemize}

\noindent \textbf{Format for Each Agent's Response:}
\begin{itemize}
    \item \textbf{Agent [Name] Analysis:}
    \begin{itemize}
        \item \textbf{Purpose Analysis:} [Describe alignment/conflict and reasons]
        \item \textbf{Action Analysis:} [Describe alignment/conflict and reasons]
        \item \textbf{Conditions/Constraints Analysis:} [Describe alignment/conflict and reasons]
    \end{itemize}
\end{itemize}

\noindent \textbf{Steps:}
\begin{enumerate}
    \item \textbf{Agent A:} Provide your analysis.
    \item \textbf{Agent B:} Provide your analysis.
    \item \textbf{Agent C:} Provide your analysis.
\end{enumerate}

\noindent \textbf{Arbiter Instructions:}
\begin{itemize}
    \item Review all three agents' analyses.
    \item Identify the strongest reasoning for each component comparison.
    \item Produce a final, synthesized analysis that captures the most accurate and insightful points raised by the agents.
\end{itemize}

\noindent \textbf{Final Output Format (Arbiter's Consolidated Analysis):}
\begin{itemize}
    \item \textbf{Purpose Analysis:} [Consolidated Analysis]
    \item \textbf{Action Analysis:} [Consolidated Analysis]
    \item \textbf{Conditions/Constraints Analysis:} [Consolidated Analysis]
\end{itemize}

\end{mdframed}
\caption{Second ToT prompt.}
\label{fig:condition_example}
\end {figure}

\begin{figure}[h!]
\begin{mdframed}[linewidth=1pt, backgroundcolor=lightergray]
\textbf{User Prompt (Third ToT):}  \\
\noindent You are a set of three reasoning agents (\textbf{Agent A}, \textbf{Agent B}, \textbf{Agent C}), followed by a final \textbf{Arbiter}. You have received a consolidated comparison analysis from the previous step.

\noindent \textbf{Consolidated Analysis:} \texttt{\{consolidated\_analysis\}}

\noindent \textbf{Instructions for Agents:}
\begin{itemize}
    \item Each agent independently determines whether the requirement overall \textbf{Conforms} or \textbf{Violates} the reference text.
    \item Consider:
    \begin{enumerate}
        \item Purpose Alignment
        \item Action Consistency
        \item Conditions/Constraints Alignment
    \end{enumerate}
\end{itemize}

\noindent \textbf{Format for Each Agent's Response:}
\begin{itemize}
    \item \textbf{Agent [Name] Final Assessment:}
    \begin{itemize}
        \item \textbf{Overall Assessment:} [Conforms | Violates]
        \item \textbf{Rationale:} [Explain reasoning based on previous consolidated analysis]
    \end{itemize}
\end{itemize}

\noindent \textbf{Steps:}
\begin{enumerate}
    \item \textbf{Agent A:} Provide your final assessment.
    \item \textbf{Agent B:} Provide your final assessment.
    \item \textbf{Agent C:} Provide your final assessment.
\end{enumerate}

\noindent \textbf{Arbiter Instructions:}
\begin{itemize}
    \item Review the three agents' assessments.
    \item Consider which assessment is best supported by the prior analyses.
    \item Produce a final determination and rationale that either adopts one agent’s perspective entirely or synthesizes them if needed.
\end{itemize}

\noindent \textbf{Final Output Format (Arbiter's Conclusion):}
\begin{itemize}
    \item \textbf{Overall Assessment:} [Conforms | Violates]
    \item \textbf{Rationale:} [Detailed explanation incorporating the best reasoning points from the agents]
\end{itemize}

\end{mdframed}
\caption{Third ToT prompt.}
\label{fig:condition_example}
\end {figure}




\end{appendices}


\end{document}